\documentclass[aps,pra,superscriptaddress,floatfix,showpacs,10pt, column,showkeys]{revtex4}
\usepackage{bbm}
\usepackage{amsfonts}
\usepackage[dvips]{graphicx}
\usepackage{amsmath,amsfonts,amssymb,graphics,graphicx,epsfig,color,times,bbm}
\usepackage{appendix}
\usepackage{subfigure}

\usepackage{lipsum}
\usepackage{graphicx}
\if CLASSOPTIONcompsoc
\usepackage[caption=false, font=normalsize, labelfont=sf, textfont=sf]{subfig}
\else
\usepackage[caption=false, font=footnotesize]{subfig}
\fi
\DeclareMathOperator{\Tr}{Tr}

\begin{document}

\title{Quantum Fisher Information With General Quantum Coherence in multi-dimensional quantum systems}

\author{Jun-Long Zhao}
\affiliation{Quantum Information Research Center, Shangrao Normal University, Shangrao 334001, China}
\affiliation{Jiangxi Province Key Laboratory of Applied Optical Technology(2024SSY03051), Shangrao Normal University, Shangrao 334001, China}

\author{Li Yu}
\affiliation{School of Physics, Hangzhou Normal University, Hangzhou 311121, China}

\author{Ming Yang}
\thanks{mingyang@ahu.edu.cn}
\affiliation{School of Physics and Optoelectronics Engineering, Anhui University, Hefei 230601, China}

\author{Chui-Ping Yang}
\thanks{yangcp@hznu.edu.cn}
\affiliation{School of Physics, Hangzhou Normal University, Hangzhou 311121, China}

\begin{abstract}
 Quantum metrology is a science about quantum measurements and it plays a key role in precision of quantum parameter estimation.
   Meanwhile, quantum coherence is an important quantum feature and quantum Fisher information (QFI)
    is an important indicator for precision of quantum parameter estimation. In this paper, we explore the relationship
     between  QFI and quantum coherence in multi-dimensional quantum systems.
      We introduce a new concept referred to as General Quantum Coherence (GQC),
      which characterizes the quantum coherence and the eigenenergies of the Hamiltonian in the interaction processes.
      GQC captures quantum nature of high-dimensional quantum states and addresses shortcomings in coherence measurement.
      Additionally, we observe a stringent square relationship between GQC and QFI.
      This finding provides a crucial guideline for improving the precision of parameter estimation.
\end{abstract}



\maketitle

\section{introduction\label{sec:1}}

  Metrology, as a science of measurements, has had an immense impact on our world.
  It provides original data in the field of scientific research. In the realm of quantum mechanics,
  this discipline is referred to as quantum metrology, playing a crucial role in the study of quantum mechanics.
  It is indispensable for enhancing the precision of estimating unknown parameters\cite{Pezze1,Giovannetti3,Demkowicz}.
  In general, the whole estimation process can be divided into three steps: the preparation of the initial probe state $\rho_{in}$;
   the evolution of the initial probe state (described by a unitary operator $U(\varphi)$ with an unknown parameter $\varphi$);
   and the detection of the probe output state $\rho(\varphi)$. When the probe state is a quantum state,
   and it is a separable state, the parameter estimation error is limited to a scaling factor of $1/\sqrt{N_m}$,
    known as the standard quantum limit \cite{Braginsky, Giovannetti1,Giovannetti2}.

  According to the quantum Cram\'{e}r-Rao bound \cite{Braubstein,Cramer,Toth1,Helstrom,Fisher,Paris},
  the reciprocal of quantum Fisher information(QFI) provides a lower bound on the variance of the
  estimator $\hat{\varphi}$ is $\delta\hat{\varphi}\geq 1/\sqrt{F_Q}$, where $\delta\hat{\varphi}$ represents
   the standard deviation and $F_Q$ is the QFI. QFI, which encapsulates the information related to quantum states,
   has applications in various domains such as quantum information procession and transmission.
   The reciprocal of the QFI establishes the lower limit on the variance
   of the estimator, in other words, a larger QFI indicates a higher precision in parameter estimation.
   Therefore, the study of the fractional determinants of QFI holds significant importance.

   To uncover the essence of quantum metrology, extensive research has been conducted on the correlation between QFI and various properties of quantum states.
   These properties encompass quantum entanglement \cite{wanfang,linan, Toth,Hyllus,Strobel}, fidelity \cite{luijing,Banchi,Modi},
   spin squeezing parameters \cite{wanfang,zhongwei,majian2}, quantum discord \cite{Fatih,yaoyao,Girolami2}, and so on.
   However, until now, a definitive relationship between these quantum properties and QFI has not been established.
   In fact, these investigations have been delved into a range of quantum correlation properties such as
   quantum entanglement \cite{Schrodinger1,Einstein}, quantum discord \cite{Zurek,Ollivier}, quantum non-locality, quantum steering \cite{Wiseman,Jones}, spin squeezing parameters \cite{Walls,majian1}, and so on.
   These properties collectively can be considered as a form of generalized quantum coherence.
   Consequently, our attention has been drawn to exploring the connection between quantum coherence and QFI.

   In quantum metrology, attention is drawn to a phenomenon wherein two quantum states exhibit maximal coherence, namely
   \{$\frac{1}{\sqrt{2}}\left(|0\rangle+|1\rangle\right)$\} and \{$\frac{1}{\sqrt{2}}\left(|0\rangle +|N\rangle\right)$\}.
   Despite sharing the same level of coherence, these states have distinct QFI values 1 and $N^2$ respectively.
   What are the reasons behind these observed phenomena?

 In order to overcome  this question and rectify limitations in existing measurement schemes,
    this paper introduces a new concept involving quantum coherence: \emph{general quantum coherence} (GQC).
    The GQC denotes that the quantum coherence is  extended, and it includes the quantum coherence and the energy levels of quantum system.
     In 2012, Fr\"{o}wis and D\"{u}r used the QFI as a measure of macroscopicity\cite{Frowis}, and it revealed that QFI
   and macroscopicity of quantum states have an intimate connection.
  In 2017, Kwon \emph{et al} proposed a measure of macroscopic coherence based
   on the degree of disturbance caused by a coarse-grained measurement \cite{Kwon}.
 Our new concept (QGC) and  macroscopic coherence (MC) \cite{Marquardt,Yadin,ZhengQ,Kwon2,ReidMD,Naseri,Levine,Gavalcanti,LeeCW} have many similarities, however the GQC and MC  come from different sources.
The MC is from combining quantum coherence and system energy levels, and the GQC is  from the study in which the quantum coherence of the probe state plays major roles.
Concurrently, we show that QFI is equivalent to the square of GQC.
    This insight prompts an exploration into the fundamental reasons why quantum metrology outperforms classical metrology.
    The results highlight a significant and straightforward relationship between QFI and GQC.

   This paper is organized as follows.  In section  \ref{sec:QFI}, QFI is introduced.  In section \ref{sec:macoh}, GQC is introduced for pure quantum state.
   In section \ref{sec:mixGQC}, GQC is extend to mixed quantum state. In section \ref{sec:experiment}, we provide an example in which GQCs and QFIs of the states $\frac{1}{\sqrt{2}}\left(|0\rangle+|1\rangle\right)$
   and $\frac{1}{\sqrt{2}}\left(|00\rangle+|11\rangle\right)$ are investigated in experiment. Section \ref{sec:conclusion} summarizes the main findings of this work.

\section{quantum fisher information \label{sec:QFI}}

In general,  a probe initial state $\rho$ undergoes a parametrization process with a parameter $\varphi$.
The QFI for the output state $\rho_{\varphi}$ can be written by \cite{Helstrom,Holevo,LiuJ}
\begin{equation}
 F_Q = \Tr\left(\rho_\varphi L^2\right),
\end{equation}
where $F_Q$ represents QFI, $L$ is called as symmetric logarithmic derivative  and it can be expressed as follows
\begin{equation}
 \frac{\partial \rho_\varphi}{\partial \varphi}=\frac{1}{2}\left(L\rho_{\varphi}+ \rho_{\varphi} L\right).
\end{equation}
If the parametrization is a unitary process described by a unitary operator $U_{\varphi}=e^{-i\hat{H}\varphi}$, the output state of the probe is given by
\begin{equation}
 \rho_{\varphi} = U_{\varphi} \rho U_{\varphi}^{\dagger}.
\end{equation}
Here, $\hat{H}$ is the  the parametrization Hamiltonian applied in the parameter estimation process.
Without loss of generality, consider that the probe is initially in an arbitrary mixed state $\rho=\sum_{i=1}^n p_i |\psi_i\rangle \langle \psi|$,
where $|\psi_i\rangle$ is an eigenstate of $\rho$ with an eigenvalue $p_i$.
The QFI can be expressed as  \cite{majian,Pezze}
\begin{equation}
F(\rho, \hat{H})=2\sum_{i \neq j}^{n}{\frac{(p_i-p_j)^2}{p_i+p_j}}| \langle \psi_i| \hat{H}|\psi_j \rangle |^2.
\label{equ:QFI}
\end{equation}
On the other hand, consider that the probe is initially in an arbitrary pure state $|\psi\rangle$. According to \cite{majian,Pezze}, the QFI is given by
\begin{eqnarray}
 F(|\psi\rangle, \hat{H}) =    4\left( \langle \psi| \hat{H} ^2|\psi \rangle - \langle \psi| \hat{H} |\psi\rangle ^2\right). \label{equ:pureQFI}
\end{eqnarray}

\section{General quantum Coherence for Pure state\label{sec:macoh}}

Quantum coherence is the most distinguished feature of quantum mechanics,
 characterizing the superposition properties of quantum states.
 As a quantity of coherence, we use the $l_1$-norm of coherence $C_{l_1}$ \cite{Baumgratz}

\begin{equation}
 C_{l_1}=\sum_{i \neq j} \left|\rho_{ij}\right|.\label{coherence}
\end{equation}

\subsection{A two-dimensional parameterization system}

For a two-dimensional parametrization system (i.e., a qubit), the parametrization process is characterized by the Hamiltonian
\begin{equation}
 \hat{H}=\textnormal{diag} \left( \lambda_0,  \lambda_1\right), \label{Hqubit}
\end{equation}
where the eigenstates (eigenvalues) of the parametrization Hamiltonian are labeled by $\{|0\rangle, |1\rangle \}$  ($\{\lambda_0, \lambda_1\}$),
and $\textnormal {diag}\left(\cdot\right)$ denotes a diagonal matrix.

If the parametrization process is described by a spin Hamiltonian, the relationship between QFI and quantum coherence is written by \cite{ZhaoJL}
\begin{equation}
 F_Q= C^2.\label{QFIcohqubit}
\end{equation}
Because the eigenvalues of the spin Hamiltonian are $\frac{1}{2}$ and $-\frac{1}{2}$, it follows from Eq.~(\ref{QFIcohqubit})
 \begin{equation}
  F_Q=C^2\cdot\left[\frac{1}{2}-(-\frac{1}{2})\right]^2,\label{QFIcohqubit2}
 \end{equation}
 with $\lambda_0=\frac{1}{2}$ and $\lambda_1=-\frac{1}{2}$.

 According to Eq.~(\ref{QFIcohqubit2}), for an arbitrary parametrization Hamiltonian $\hat{H}$ given in Eq.~(\ref{Hqubit}),
we can \emph{infer} the following relationship
 \begin{equation}
  F_Q=C^2\cdot\left(\lambda_0-\lambda_1\right)^2.\label{QFIcohqubit3}
 \end{equation}

 In order to prove Eq.~(\ref{QFIcohqubit3}), we first consider that the probe initial state is an arbitrary pure state  given by
  \begin{equation}
   |\psi\rangle=c_k | k\rangle + c_l |l\rangle,\label{qubitM}
  \end{equation}
  where $|k\rangle$ and $|l\rangle$ are the two basis states, which are chosen as the two eigenstates of the parametrization Hamiltonian $\hat{H}$ with eigenvalues $\lambda_k$ and $\lambda_l$, which are given by
\begin{eqnarray}
\lambda_k & = & \langle k|\hat{H}|k\rangle,\nonumber\\
\lambda_l & = & \langle l|\hat{H}|l\rangle.  \label{eq12}
\end{eqnarray}
Thus, according to Eq.~(\ref{coherence}), the coherence of the state in  Eq.~(\ref{qubitM}) is
\begin{equation}
 C\left(|\psi\rangle\right)=2|c_k c_l^*|. \label{qubitMcoh}
\end{equation}
Note that for a pure state $|\psi\rangle$, the QFI is expressed as Eq.~(\ref{equ:pureQFI}).
Thus, base on Eq.~(\ref{equ:pureQFI}) and Eq.~(\ref{eq12}), one can easily find that the QFI for the state in Eq.~(\ref{qubitM}) is
 \begin{eqnarray}
   F_Q\left(|\psi\rangle, \hat{H}\right)   & = & 4\left[ \langle \psi| \hat{H} ^2|\psi \rangle - \left(\langle \psi| \hat{H} |\psi\rangle \right)^2\right]\nonumber\\
                                                                & = & 4\left[|c_k|^2\lambda_k^2+|c_l|^2\lambda_l^2- (|c_k|^2\lambda_k+|c_l|^2\lambda_l)^2\right]\nonumber\\
                                                                & = & 4\left[|c_k|^2(1-|c_k|^2)\lambda_k^2+|c_l|^2(1-|c_l|^2)\lambda_l^2- 2|c_k|^2|c_l|^2 \lambda_k \lambda_l \right]\nonumber\\
                                                               & = & 4|c_k c_l^*|^2 \cdot (\lambda_k - \lambda_l)^2\nonumber\\
                                                                & = & C^2 \cdot  (\lambda_k - \lambda_l)^2,   \label{qubitQFI}
 \end{eqnarray}
 which shows that the QFI  is determined by the quantum coherence $C$ of the probe initial state $|\psi\rangle$ and
 the difference $\lambda_k-\lambda_l$ between the eigenvalues of the parametrization Hamiltonian corresponding to the basis states.

 From Eq.~(\ref{qubitQFI}),  $C\left(|\psi\rangle\right) \cdot \left|\lambda_i-\lambda_j\right|$ is defined as a new quantity as follows
  \begin{equation}
   \mathcal{M}=C \cdot \left|\lambda_i -\lambda_j \right|.\label{qubitPGQC}
  \end{equation}
  Here, $\mathcal{M}$ is the extension of quantum coherence, it includes the coherence $C$ and the energy level difference $|\lambda_i - \lambda_j|$ of two basis states.
  Therefore, we call this new concept as general quantum coherence (GQC).

  \subsection{A high-dimensional parametrization system}
  In this subsection, we will extend the new conception (\ref{qubitPGQC})
to a high-dimensional parametrization system (i.e., a qudit). For an $n$-dimensional system, the parametrization process is characterized by the Hamiltonian
\begin{equation}
 \hat{H}=\textnormal{diag} \left(\lambda_0, \lambda_1, \lambda_2, \cdots, \lambda_{n-1}\right), \label{Hqudit}
 \end{equation}
 where $\lambda_0, \lambda_1, \lambda_1, \cdots, ~\textnormal{and}~\lambda_{n-1}$ are the eigenvalues of the parametrization Hamiltonian $\hat{H}$  and the corresponding eigenstates of the parametrization Hamiltonian $\hat{H}$  are labeled by  $|0\rangle, |1\rangle, \cdots$, and $|n-1\rangle$. The eigenvalues are given by
 \begin{equation}
  \lambda_i=\langle i|\hat{H}|i\rangle, \label{Hquditvalue}
 \end{equation}
 where $i=0, 1, 2, \cdots, n-1$. For such a system, an initial arbitrary pure state of the probe can be written as
 \begin{equation}
  |\psi\rangle=a_0|0\rangle+a_1|1\rangle+...+a_{n-1}|n-1\rangle,\label{equ:purqudit}
 \end{equation}
 where the basis states $|0\rangle$, $|1\rangle$, ..., and $|n-1\rangle$ are chosen as the eigenstates of the parametrization Hamiltonian $\hat{H}$.

   In order to quantify coherence of the state $|\psi\rangle$,
   the $n$-level quantum system will be divided into many two-level subsystems.
   For the subsystem with two levels $|k\rangle$ and $|l\rangle$, an arbitrary quantum pure state is given by
   \begin{equation}
     |\psi_{kl}\rangle=a_k|k\rangle+a_l|l\rangle, \label{equ:purqubit}
   \end{equation}
   where $k, l=0, 1, 2, ..., n-1$. So, the coherence of the state $|\psi_{kl}\rangle$ is  $2|a_k a_l^*|$,
   corresponding to the GQC being  $2|a_k a_l^*|\cdot \left|\lambda_k-\lambda_l\right|$.
   Considering all the cases, the possible definitions of the GQC of the pure state $|\psi\rangle$ in Eq.~(\ref{equ:purqudit}) are written as:
  \begin{equation}
    \mathcal{M}\left(|\psi\rangle\right) = \sqrt{\sum_{i<j}4\left|a_i a_j^*\right|^2(\lambda_i-\lambda_j)^2}\label{pureGQC1},
  \end{equation}
   or
  \begin{equation}
    \mathcal{M}\left(|\psi\rangle,H\right) = \sum_{i<j} 2 \left|a_i a_j^*\right|\left|\lambda_i-\lambda_j\right|\label{pureGQC2}.
  \end{equation}

  From Eq.~(\ref{pureGQC1}) and Eq.~(\ref{pureGQC2}), we think that Eq.~(\ref{pureGQC1}) is more appropriate
  than  Eq.~(\ref{pureGQC2}), because the quantum coherence is a vector superposition
  and should satisfy the vector superposition principle.  We use Eq.~(\ref{pureGQC1}) as
  the definition of GQC. The QFI of Eq.~(\ref{equ:purqudit}) is
  \begin{eqnarray}
   F_Q \left(|\psi\rangle,H\right) & = & 4\left(\langle H^2 \rangle - \langle H\rangle^2 \right)\nonumber\\
       & = & \sum_{i<j}^{n-1} 4 |a_i|^2 |a_j|^2 \left(\lambda_i-\lambda_j\right)^2.\label{pureQFI}
  \end{eqnarray}
  Base on Eq.~(\ref{pureGQC1}) and Eq.~(\ref{pureQFI}), the relationship between GQC and QFI is
  \begin{equation}
   F_Q \left(|\psi\rangle,H\right) = \mathcal{M}^2\left(|\psi\rangle\right).\label{pureGQCQFI}
  \end{equation}
  From Eq.~(\ref{pureGQCQFI}), we find that QFI of a pure quantum state is equal to square of GQC.

  \section{General quantum Coherence for mixed state }\label{sec:mixGQC}

  In the previous section, we have defined the GQC for pure state and the determining factors of QFI when the probe initial state is a pure state. In this section, we will extend the GQC to mixed state and discuss the case when the probe initial state is a mixed state.

  \subsection{A two-dimensional parametrization system}

  In order to study the mixed-state case, we first consider a two-dimensional parametrization system.
 In this case,  an initial arbitrary mixed state of the probe can be written as
  \begin{equation}
   \rho=p_0 |0_{\vec{n}}\rangle \langle 0_{\vec{n}}| +p_1 |1_{\vec{n}}\rangle \langle 1_{\vec{n}}|, \label{equ:mixqubit}
  \end{equation}
  where $p_0+p_1=1$, $p_0$ and $p_1$ are two eigenvalues of $\rho$,  $|0_{\vec{n}}\rangle$ and $|1_{\vec{n}}\rangle$ are two eigenstates of $\rho$. We suppose that the two eigenstates are
  \begin{equation}
   \left\{
           \begin{array}{c}
             |0_{\vec{n}}\rangle = a_0 |0\rangle + a_1 |1\rangle, \\
             |1_{\vec{n}}\rangle = b_0 |0\rangle + b_1|1\rangle,
           \end{array}
         \right.
  \end{equation}
  where $|0\rangle$ and $|1\rangle$ are the eigenstates of the parametrization Hamiltonian $\hat{H}$ given in Eq.~(\ref{Hqubit}) with eigenvalues $\lambda_0=\langle 0|\hat{H}|0\rangle$ and $\lambda_1=\langle1|\hat{H}|1\rangle$ respectively.
Because of $\langle 0_{\vec{n}}| 1_{\vec{n}}\rangle=0$ and $|0_{\vec{n}}\rangle \langle 0_{\vec{n}}|+|1_{\vec{n}}\rangle \langle 1_{\vec{n}}|=\hat{I}$, one has
   \begin{equation}
    \left\{
           \begin{array}{c}
              a_0^* a_1 + b_0^* b_1 =0,\\
              a_0 a_1^* +b_0 b_1^* =0,\\
              a_0 b_0^* + a_1 b_1^*=0.
           \end{array}
         \right.
  \end{equation}
  Thus, the square of coherence of the probe initial state $\rho$ given in Eq.~(\ref{equ:mixqubit}) is
  \begin{eqnarray}
   C^2\left(\rho\right) & = & 4\left|p_0 a_0 a_1^* + p_1 b_0 b_1^*\right|^2\nonumber\\
                            & = & 4\left|p_0 a_0 a_1^* - p_1 a_0 a_1^*\right|^2\nonumber\\
                            & = & 4 (p_0 - p_1)^2 \left|a_0 a_1^* \right|^2\nonumber\\
                            & = & 4 (p_0 - p_1)^2 a_0 a_1^* a_0^* a_1 \nonumber\\
                        & = & -2\left(p_0-p_1\right)^2 \left(a_0 a_1^* b_0^* b_1 +a_0^* a_1 b_0 b_1^*\right).\label{qubitCoh}
  \end{eqnarray}
According to Eq.~(\ref{equ:QFI}), the QFI is
 \begin{eqnarray}
  F_Q(\rho, \hat{H}) & = & \frac{4(p_0-p_1)^2}{p_0+p_1}\left|\langle 0_{\vec{n}}|\hat{H}| 1_{\vec{n}}\rangle\right|^2\nonumber\\
                           & = & 4(p_0-p_1)^2 \left|\langle 0_{\vec{n}}|\hat{H}| 1_{\vec{n}}\rangle\right|\cdot \left|\langle 1_{\vec{n}}|\hat{H}| 0_{\vec{n}}\rangle\right|\nonumber\\
                           & = & 4(p_0-p_1)^2 \left(a_0^*b_0\lambda_0 +a_1^*b_1\lambda_1\right)\left(a_0 b_0^* \lambda_0 +a_1 b_1^*\lambda_1 \right)\nonumber \\
                           & = & 4(p_0-p_1)^2 \left(a_0^*b_0 a_0 b_0^* \lambda_0^2 +a_1^*b_1 a_1 b_1^* \lambda_1^2 + a_1^*b_1 a_0 b_0^* \lambda_0 \lambda_1+a_0^*b_0 a_1 b_1^*  \lambda_0\lambda_1 \right)\nonumber\\
                           & = & -2\left(p_0-p_1\right)^2 \left(a_0 a_1^* b_0^* b_1 +a_0^* a_1 b_0 b_1^*\right)\left(\lambda_0-\lambda_1\right)^2\nonumber\\
                           & = & C^2(\rho) \left(\lambda_0-\lambda_1\right)^2.\label{MixqubitQFI}
 \end{eqnarray}
 Eq.~(\ref{MixqubitQFI}) shows that the QFI is determined by the quantum coherence $C(\rho)$ of the probe initial state $\rho$
  and the difference $\lambda_0-\lambda_1$ between the eigenvalues of the parametrization Hamiltonian corresponding to the basis states.
 Thus, according to Eq.~(\ref{qubitPGQC}), the square of GQC is defined by
 \begin{equation}
  \mathcal{M}^2\left(\rho\right) = -2\left(p_0-p_1\right)^2 \left(a_0 a_1^* b_0^* b_1 +a_0^* a_1 b_0 b_1^*\right)\left(\lambda_0-\lambda_1\right)^2.\label{qubitGQC}
 \end{equation}
 Thus, we have the relationship between GQC and QFI
 \begin{equation}
  F_Q(\rho, \hat{H}) =\mathcal{M}^2\left(\rho\right).
 \end{equation}

 \subsection{A high-dimensional parametrization system}

 Now we extend the GQC (\ref{qubitGQC}) to a high-dimensional parametrization system.
In the following, we will consider an $n$-dimensional system. The parametrization process is
characterized by the Hamiltonian (\ref{Hqudit}) with eigenvalues given in Eq.~(\ref{Hquditvalue}).
An initial arbitrary mixed state of the probe can be expressed as
 \begin{equation}
  \rho=\sum_{i=1}^n p_i |\psi_i\rangle \langle \psi_i |, \label{qudit}
 \end{equation}
  where $|\psi_i\rangle$ is an eigenstate of $\rho$ given in Eq.~({\ref{qudit}}), $p_i$ is an eigenvalue of  $\rho$ (i.e., the probability of $|\psi_i\rangle$ appearing in the mixed state $\rho$), and $\sum_{i=1}^n p_i=1$.
The eigenstates of $\rho$ are expressed as
 \begin{eqnarray}
       p_1 : & |\psi_1\rangle = a_0^{(1)} |0\rangle + a_1^{(1)}|1\rangle + a_2^{(1)}|2\rangle + \cdots + a_{n-1}^{(1)}|n-1\rangle,\nonumber\\
       p_2 : & |\psi_2\rangle = a_0^{(2)} |0\rangle + a_1^{(2)}|1\rangle + a_2^{(2)}|2\rangle + \cdots + a_{n-1}^{(2)}|n-1\rangle,\nonumber\\
             &      \vdots \nonumber\\
      p_i : & |\psi_i\rangle = a_0^{(i)} |0\rangle + a_1^{(i)}|1\rangle + a_2^{(i)}|2\rangle + \cdots + a_{n-1}^{(i)}|n-1\rangle,\nonumber\\
              &    \vdots \nonumber\\
       p_j : & |\psi_j\rangle = a_0^{(j)} |0\rangle + a_1^{(j)}|1\rangle + a_2^{(j)}|2\rangle + \cdots + a_{n-1}^{(j)}|n-1\rangle,\nonumber\\
                 &    \vdots \nonumber\\
       p_n : & |\psi_j\rangle = a_0^{(n)} |0\rangle + a_1^{(n)}|1\rangle + a_2^{(n)}|2\rangle + \cdots + a_{n-1}^{(n)}|n-1\rangle.\label{eigenpsi}
  \end{eqnarray}
where the basis states $|0\rangle$, $|1\rangle$,  $\cdots$, and $|n-1\rangle$ are chosen as eigenstates of the parametrization Hamiltonian $\hat{H}$ with eigenvalues $\lambda_0$, $\lambda_1$, $\cdots$, and $\lambda_{n-1}$ respectively. Here, $\lambda_i=\langle i|\hat{H}|i\rangle$ ($i=0, 1, \cdots, n-1$).

 For the eigenstates of $\rho$ given in Eq.~(\ref{eigenpsi}), we take two arbitrary eigenstates ($|\psi_i\rangle$, $|\psi_j\rangle$) to form a new quantum state
 \begin{equation}
  \widetilde{\rho}_{i,j}=\frac{1}{p_i+p_j}(p_i |\psi_i\rangle \langle \psi_i |+p_j | \psi_j \rangle \langle \psi_j |),\label{equ:subqudit}
 \end{equation}
 which is a part of the mixed state $\rho$.
 The two  eigenstates $|\psi_i\rangle$ and $|\psi_j\rangle$ are given by
  \begin{eqnarray}
   |\psi_i\rangle = a_0^{(i)} |0\rangle + a_1^{(i)}|1\rangle + a_2^{(i)}|2\rangle + \cdots + a_{n-1}^{(i)}|n-1\rangle,\\
   |\psi_j\rangle = a_0^{(j)} |0\rangle + a_1^{(j)}|1\rangle + a_2^{(j)}|2\rangle + \cdots + a_{n-1}^{(j)}|n-1\rangle.
 \end{eqnarray}
 Now, we choose two arbitrary basis states $\left\{|k\rangle, |l\rangle\right\}$ to construct the following two pure states of a quantum state of a two-dimensional subsystem
  \begin{eqnarray}
  |\psi_i^{(kl)}\rangle=a_k^{(i)}|k\rangle + a_l^{(i)}|l\rangle, \\
   |\psi_j^{(kl)}\rangle=a_k^{(j)}|k\rangle + a_l^{(j)}|l\rangle.
 \end{eqnarray}
 In this two-dimensional subsystem, we adopt $|\psi_i^{(kl)}\rangle$ and  $|\psi_j^{(kl)}\rangle$ to form the  following quantum state
  \begin{equation}
    \widetilde{\rho}_{i,j}^{(kl)}=\frac{1}{p_i+p_j}\left(p_i |\psi_i^{(kl)}\rangle \langle \psi_i^{(kl)} |+p_j | \psi_j^{(kl)} \rangle \langle \psi_j^{(kl)} |\right).\label{equ:subqubit}
  \end{equation}

\begin{figure}[!htbp]
\setlength{\belowcaptionskip}{0cm}
       \includegraphics[scale=1]{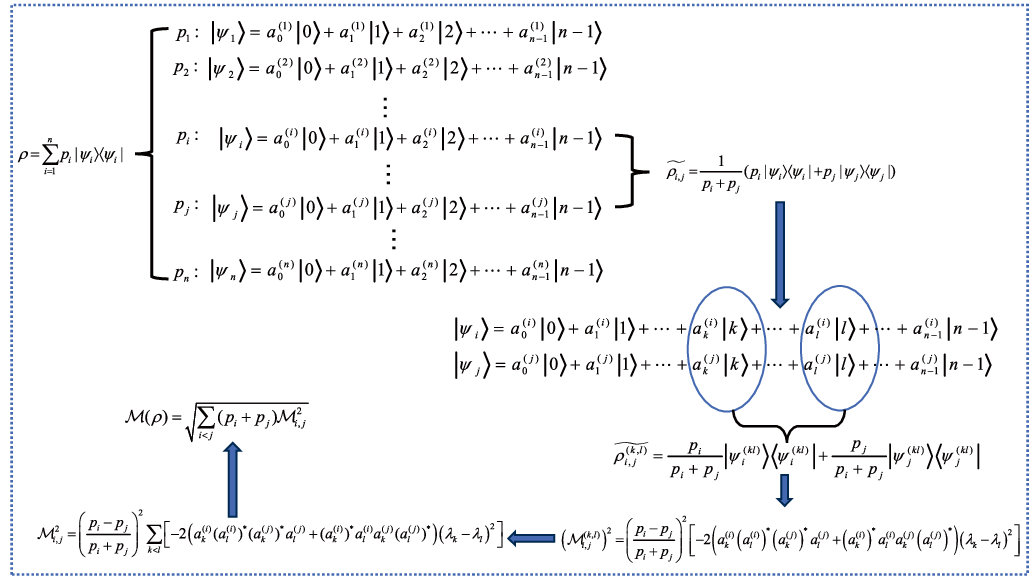}\hspace*{0pt}
\caption{The illustration of calculation process  for mixed state case. }\label{fig3}
\end{figure}

 According to Eq.~(\ref{qubitCoh}), the coherence of this constructed quantum state $\widetilde{\rho}_{i,j}^{(kl)}$ is
  \begin{equation}
  \left(C_{i,j}^{(kl)}\right)^2 =-2 \left(\frac{p_i-p_j}{p_i+p_j}\right)^2 \left[a_k^{(i)} a_l^{* (i)} a_k^{*(j)} a_l^{(j)} + a_k^{*(i)} a_l^{(i)} a_k^{(j)} a_l^{*(j)}\right],
\end{equation}
and the GQC $\mathcal{M}_{i,j}^{(kl)}$ of Eq.~(\ref{equ:subqubit})  can be written as
\begin{equation}
  \left(\mathcal{M}_{i,j}^{(kl)}\right)^2 =-2 \left(\frac{p_i-p_j}{p_i+p_j}\right)^2 \left[a_k^{(i)} a_l^{*(i)} a_k^{*(j)} a_l^{(j)} +a_k^{*(i)} a_l^{(i)} a_k^{(j)} a_l^{*(j)}\right]\left(\lambda_k-\lambda_l\right)^2.
\end{equation}
For the constructed quantum state $ \widetilde{\rho}_{i,j}$  in Eq.~(\ref{equ:subqudit}), we thus obtain
  \begin{equation}
  \mathcal{M}_{i,j}^2 = \left(\frac{p_i-p_j}{p_i+p_j}\right)^2 \sum_{k<l} \left[-2\left(a_k^{(i)} a_l^{*(i)} a_k^{*(j)} a_l^{(j)} + a_k^{*(i)} a_l^{(i)} a_k^{(j)} a_l^{*(j)}\right)\left(\lambda_k-\lambda_l\right)^2 \right].\label{eqa:mutuGQC}
 \end{equation}
Therefore, for the probe initial mixed state  $\rho$ in Eq.~(\ref{qudit}), the GQC of $\rho$ given in Eq.~(\ref{qudit}) can be written as follows
 \begin{equation}
 \mathcal{M}(\rho)=\sqrt{\sum_{i<j}(p_i+p_j)\mathcal{M}_{i,j}^2}.\label{eqa:mixGQC}
 \end{equation}
 The calculation process for obtaining the expression of $ \mathcal{M}(\rho)$ given in Eq.~(\ref{eqa:mixGQC}) is illustrated in Fig. \ref{fig3}.

The QFI for a mixed state $\rho$ in Eq.~(\ref{qudit}) is expressed as Eq.~(\ref{equ:QFI}). We calculate  $| \langle \psi_i| H |\psi_j \rangle |^2$ and obtain
  \begin{equation}
   | \langle \psi_i| \hat{H} |\psi_j \rangle |^2 = -\frac{1}{4}\sum_{k\neq l}^{n-1}\left[a_k^{(i)} a_l^{*(i)} a_k^{*(j)} a_l^{(j)} +a_k^{*(i)} a_l^{(i)} a_k^{(j)} a_l^{*(j)}\right](\lambda_k-\lambda_l)^2.
  \end{equation}
According to Eq.~(\ref{equ:QFI}), we thus have
  \begin{eqnarray}
   F_Q\left(\rho, \hat{H}\right) & = & -\frac{1}{2}\sum_{i\neq j}^{n}\frac{\left(p_i-p_j\right)^2}{p_i+p_j}  \sum_{k\neq l} ^{n-1}\left[a_k^{(i)} a_l^{*(i)} a_k^{*(j)} a_l^{(j)} +a_k^{*(i)} a_l^{(i)} a_k^{(j)} a_l^{*(j)}\right]\left(\lambda_k-\lambda_l\right)^2\nonumber\\
    & = & \sum_{i< j}^{n}\frac{\left(p_i-p_j\right)^2}{p_i+p_j}  \sum_{k< l} ^{n-1}\left\{-2\left[a_k^{(i)} a_l^{*(i)} a_k^{*(j)} a_l^{(j)} +a_k^{*(i)} a_l^{(i)} a_k^{(j)} a_l^{*(j)}\right]\left(\lambda_k-\lambda_l\right)^2 \right\}\nonumber\\
    & = &  \sum_{i< j}^{n} (p_i+p_j) \sum_{k< l} ^{n-1} \left\{-2\left(\frac{p_i-p_j}{p_i+p_j}\right)^2\left[a_k^{(i)} a_l^{*(i)} a_k^{*(j)} a_l^{(j)} +a_k^{*(i)} a_l^{(i)} a_k^{(j)} a_l^{*(j)}\right]\left(\lambda_k-\lambda_l\right)^2 \right\} \nonumber\\
    & = & \sum_{i<j}(p_i+p_j)\mathcal{M}_{i,j}^2\nonumber\\
     & = &  \mathcal{M}^2(\rho)\label{mixquditQFI}
  \end{eqnarray}

  Based on Eq.~(\ref{mixquditQFI}), the relationship between QFI and GQC can be written as follows
  \begin{equation}
    F_Q(\rho,H)=\mathcal{M}^2(\rho).\label{QFIGQC}
  \end{equation}
 Based on Eq.~(\ref{QFIGQC}), we find that QFI of parameter estimation is equal to the square of GQC of the probe state.
 This result implies that the QFI is determined by the GQC. The GQC can be enhanced by increasing the GQC.

\section{Experiment \label{sec:experiment}}

\begin{figure}[!htbp]
\setlength{\belowcaptionskip}{0cm}
       \includegraphics[scale=0.5]{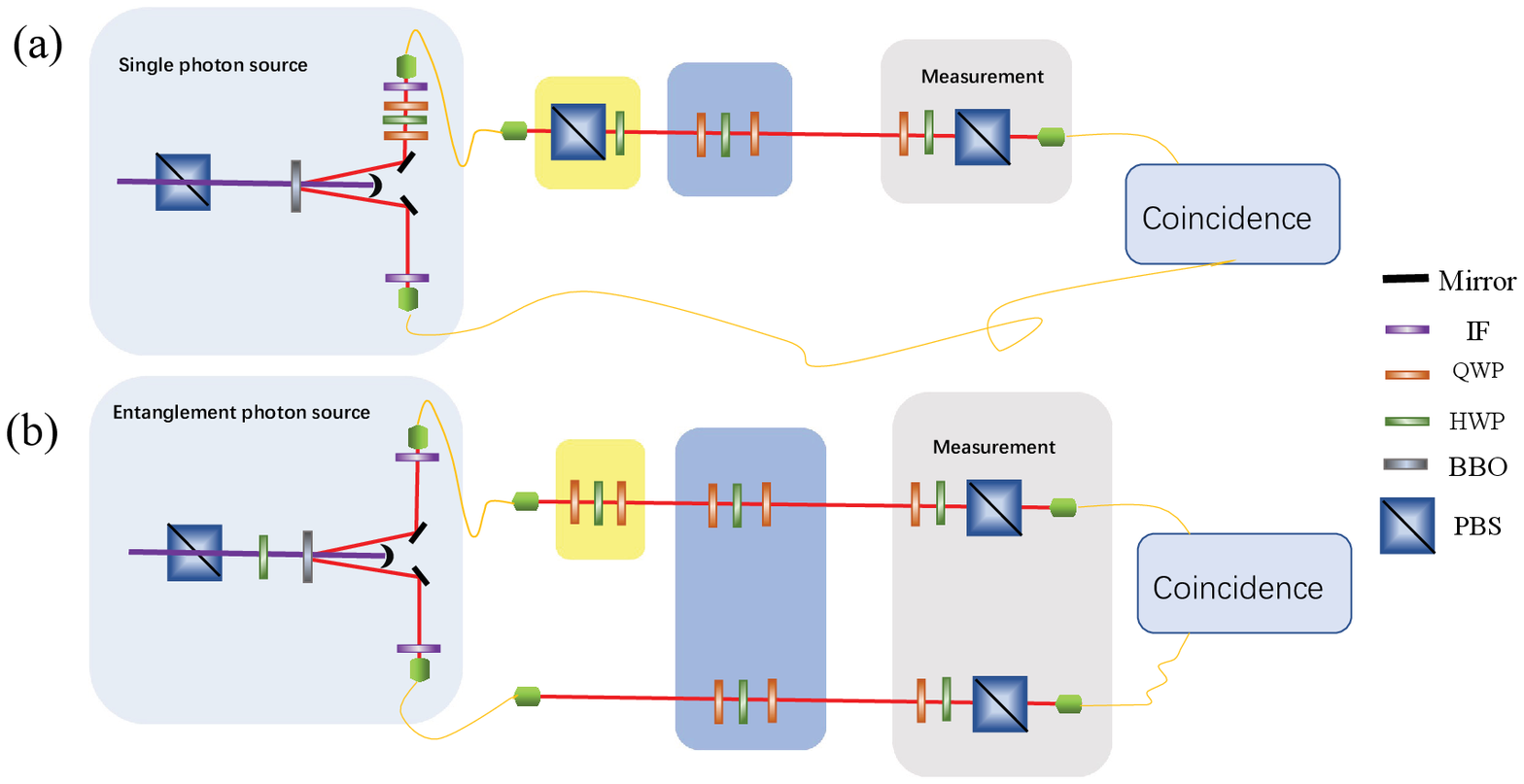}\hspace*{130pt}
\caption{\textbf{Experiment setup:}  (a) Quantum parameter estimation process when the parametrization system is a  single qubit system.
The single photons source is created through a degenerated spontaneous
parametric down-conversion (SPDC) process by pumping a type-I phase nonlinear $\beta$-barium-borate (BBO) crystal in the aqua area.
The preparation of the probe state is shown in the yellow area.
 The parametrization process is shown in the blue area.
 The measurement of the output state is shown  in the gray area.
(b) Quantum parameter estimation process when the parametrization system is a  two-qubit system.
A pair of entangled photons are generated through degenerated  (SPDC) process by pumping two type-II phase  perpendicular each other nonlinear $\beta$-BBO
crystals in the aqua area.
 The parametrization process is shown in the  blue area.
 The measurement of output state is shown in the gray area.
 In the yellow area, the target quantum state is prepared by the local operation on the entangled photon source with a maximum entangled state.
 IF: Interference filter,
 QWP: quarter-wave plate,
 HWP: half-wave plate,
 PBS: polarization beam splitter.
 }\label{fig1}
\end{figure}

To confirm the above results, we design an experiment in a linear optics system. The experiment setup is shown in Fig. \ref{fig1}. We choose the following two states as the probe initial states
 \begin{equation}
  |\psi\rangle_1 = \frac{1}{\sqrt{2}} \left( |0\rangle + |1\rangle\right), \label{Eq:psi1}
 \end{equation}
and
 \begin{equation}
  |\psi\rangle_2 = \frac{1}{\sqrt{2}} \left( |00\rangle + |11\rangle\right).\label{Eq:psi2}
 \end{equation}
The parametrization Hamiltonians are respectively given by
 \begin{equation}
 \hat{H}_1=\frac{1}{2}\sigma_z=\frac{1}{2}\left(
                                      \begin{array}{cc}
                                        1 & 0 \\
                                        0 & -1 \\
                                      \end{array}
                                    \right),
 \end{equation}
  and
  \begin{equation}
 \hat{H}_2=\frac{1}{2}\left(\sigma_z \otimes \hat{I} + \hat{I} \otimes \sigma_z\right)=\left(
                                                                                      \begin{array}{cccc}
                                                                                        1& 0 & 0 & 0 \\
                                                                                        0 & 0 & 0 & 0 \\
                                                                                       0 & 0 & 0 & 0 \\
                                                                                        0  & 0 &  0 & -1 \\
                                                                                      \end{array}
                                                                                    \right).
 \end{equation}

In experiment, we use the horizontal polarization state $|H\rangle$ to denote $|0\rangle$ and the vertical polarization state $|V\rangle$ to denote $|1\rangle$.
Thus, the two states mentioned above can be written as
  \begin{eqnarray}
  |\psi\rangle_1  & = &  \frac{1}{\sqrt{2}} \left( |H\rangle + |V\rangle\right),\\
  |\psi\rangle_2  & = &  \frac{1}{\sqrt{2}} \left( |HH\rangle + |VV\rangle\right).
 \end{eqnarray}

\begin{figure}[!htbp]
 \centering
	  \subfigure{
       \includegraphics[width=0.35\linewidth]{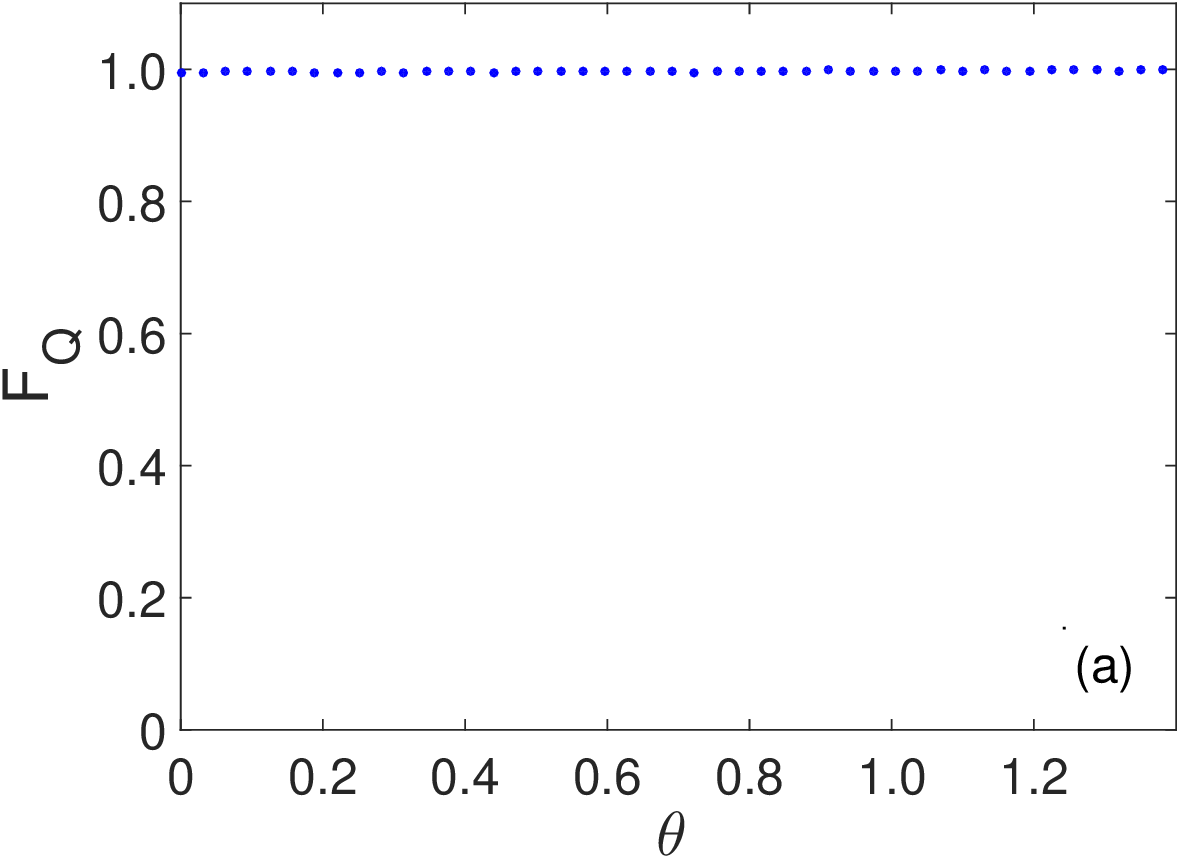}}\hspace{10mm}
    \label{1a}
	  \subfigure{
        \includegraphics[width=0.35\linewidth]{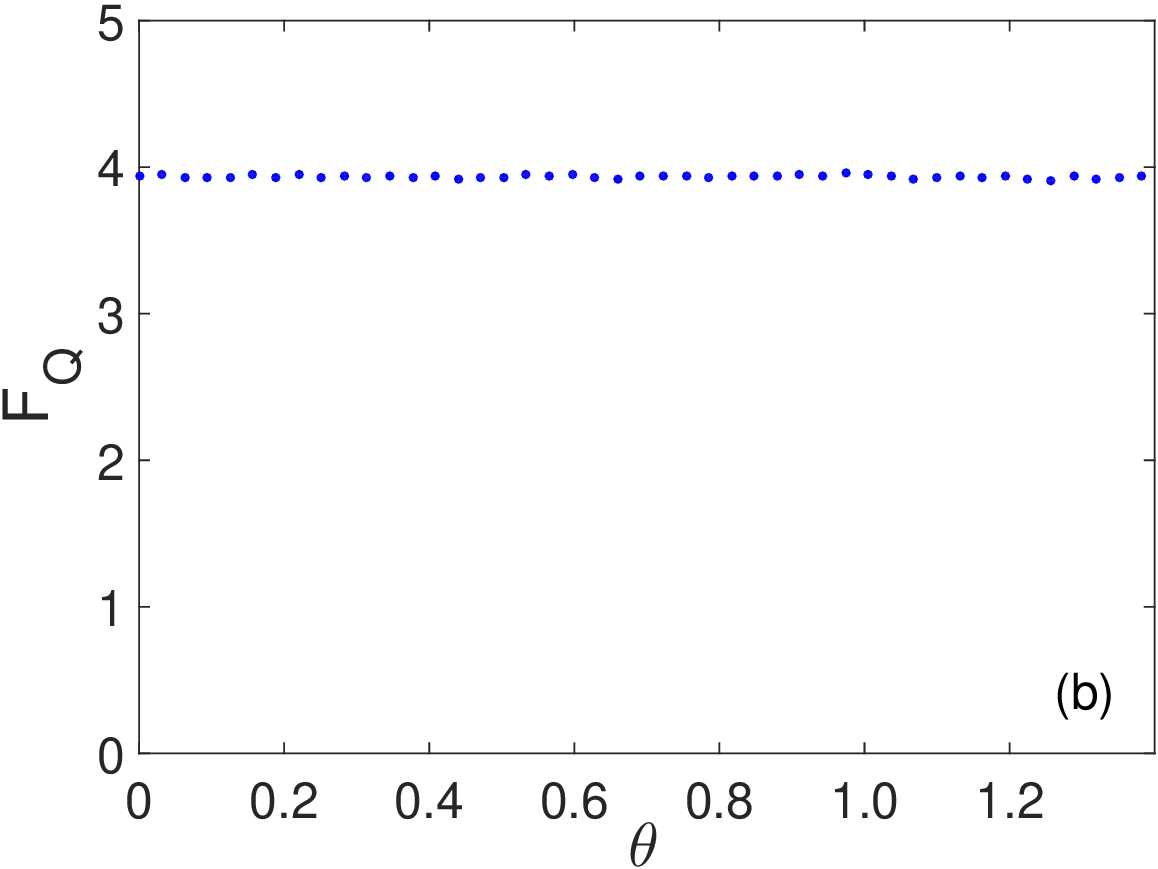}}\hspace{10mm}
    \label{1b}
\caption{The QFI varies with the estimated parameter $\theta$. The subfigures (a) denotes the QFI results when $|\psi\rangle_1$ undergoes an evolution under $\hat{H}_1$,  while (b) denotes the QFI results when $|\psi\rangle_2$ undergoes an evolution under $\hat{H}_2$. }
\label{fig2}
\end{figure}

To produce the two quantum states $|\psi_1\rangle$ and $|\psi_2\rangle$, we use the single photon source and the entanglement photon source, respectively.
 We use one barium borate (BBO) crystal to realize the spontaneous parametric down-conversion (SPDC),
 and then use a polarization beam splitter (PBS) and a half wave plate (HWP) with $22.5^{\circ}$ to produce
 the state $|\psi\rangle_1=\frac{1}{\sqrt{2}}\left( |H\rangle + |V\rangle\right)$ by the single photon [Fig. \ref{fig1}(a)].
 We use two BBO crystals (perpendicular each other) to realize SPDC and produce a pair of entangled photons.
 We first adjust the polarization of the pump light by a HWP with $22.5^{\circ}$
 and then let the pump light pass through two BBO crystals (perpendicular each other)  to
 produce the maximally entangled state $|\psi\rangle_2=\frac{1}{\sqrt{2}}\left( |HH\rangle + |VV\rangle\right)$[Fig. \ref{fig1}(b)] .
 The experimental results show that:
 (I) The average QFI for $|\psi\rangle_1$ ($F_{Q_1}$) is 0.9973 [Fig. \ref{fig2}(a)].
 (II) The average  QFI for $|\psi\rangle_2$ ($F_{Q_2}$) is 3.9346 [Fig. \ref{fig2}(b)].
 Thus, we obtain
 \begin{equation}
  F_{Q_2} \approx 4 F_{Q_1}.\label{Eq:exp}
 \end{equation}
 We note that the coherence of $|\psi\rangle_1$ in Eq.~(\ref{Eq:psi1}) and $|\psi\rangle_2$ in Eq.~(\ref{Eq:psi2}) are both 1.
 The difference between the eigenvalues of the Hamiltonian $H_1$ for the two basis state $|0\rangle$ and $|1\rangle$ involved in the state $|\psi\rangle_1$ is 1.
 The difference between the eigenvalues of the Hamiltonian $H_2$ for the two basis states $|00\rangle$ and $|11\rangle$ involved in the state $|\psi\rangle_2$ is 2.
 Thus, based on Eq.~(\ref{qubitQFI}), we have $F_{Q_2}=4F_{Q_1}$ in theory, which is confirmed by the experiment result given in Eq.~(\ref{Eq:exp}).

\section{Conclusion\label{sec:conclusion}}
We find that quantum coherence is the intrinsic reason why quantum metrology
 can make a better performance than classical metrology.
Quantum coherence of the probe states plays a special role in quantum metrology, and this is in contrast
 to the case of classical metrology.
 However, the quantum coherence has a limitation because it is not related to the
 difference of energy levels of two basis states. In order to overcome this defect,
 the quantum coherence is extended and general quantum coherence (GQC) is introduced.
 We have studied the relationship between GQC and QFI, which can be expressed by a simple formula Eq.~(\ref{QFIGQC}).
 The result shows that the GQC of the probe states plays a crucial role in the quantum parameter estimation.
 Our findings open a new avenue to improve the precision of quantum parameter estimation in the future.

\begin{acknowledgments}
This work was partly supported by the National Natural Science Foundation of China (NSFC) (Grant Nos. 12204311, 11974096, U21A20436),
the Jiangxi Natural Science Foundation (Grant No. 20224BAB211025), and
the Natural Science Foundation of Anhui Province (Grant Nos. 2008085MA16 and 2008085QA26).
\end{acknowledgments}

\end{document}